\documentclass[preprint]{aastex}

\usepackage{afterpage}

\newcommand{\fe}{[Fe/H]}

\newcommand{\degsq}{deg$^2$}

\newcommand{\mgb}{Mg~$b$}

\def\mt2{$M\!-\!T_2$}
\def\m51{$M\!-\!51$}

\def\bv{$B\!-\!V$}
\newcommand{\vi}{$V\!-\!I$}

\tabletypesize{\scriptsize}

\shorttitle{Morrison et al.}
\shortauthors{Searching for Halo Giants}

\begin{document}


\title{Mapping the Galactic Halo IV. Finding Distant Giants
Reliably with the Washington System}

\author{Heather L. Morrison\altaffilmark{1,2}}
\affil{Department of Astronomy, 
Case Western Reserve University, Cleveland OH 44106-7215 
\\ electronic mail: heather@vegemite.astr.cwru.edu}
\altaffiltext{1}{Cottrell Scholar of Research Corporation and NSF CAREER
fellow}
\altaffiltext{2}{and Department of Physics}
\author{Edward W. Olszewski}
\affil{Steward Observatory, University of Arizona, Tucson, Arizona 85726
\\ electronic mail: edo@as.arizona.edu}
\author{Mario Mateo}
\affil{Department of Astronomy, University of Michigan, 821
Dennison Bldg., Ann Arbor, MI 48109--1090\\
electronic mail: mateo@astro.lsa.umich.edu} 
\author{John E. Norris}
\affil{Mount Stromlo and Siding Spring Observatories, ANU, Private Bag,
Weston Creek PO, 2611 Canberra, ACT, Australia\\
electronic mail: jen@mso.anu.edu.au}
\author{Paul Harding}
\affil{Steward Observatory, University of Arizona, Tucson, Arizona 85726
\\ electronic mail: harding@billabong.astr.cwru.edu}
\author{R.C. Dohm-Palmer\footnote{Postdoctoral Research Scientist,
Steward Observatory}}
\affil{Department of Astronomy, University of Michigan, 821
Dennison Bldg., Ann Arbor, MI 48109--1090\\
electronic mail: rdpalmer@astro.lsa.umich.edu}
\author{Kenneth C. Freeman}
\affil{Mount Stromlo and Siding Spring Observatories, ANU, Private Bag,
Weston Creek PO, 2611 Canberra, ACT, Australia\\
electronic mail: kcf@mso.anu.edu.au}
\eject


\begin{abstract}

We critically examine the use of the Washington photometric system
(with the 51 filter) for identifying distant halo giants.  While this
is the most powerful photometric technique for isolating G and K giant
stars, spectroscopic follow-up of giant candidates is vital.  There are
two situations in which interlopers outnumber genuine giants in the
diagnostic \m51/\mt2\ plot, and are indistinguishable photometrically
from the giants. (1) In deep surveys covering tens of square degrees,
very metal-poor halo dwarfs are a significant contaminant.  An example
is our survey of the outer halo \citep{spag1,spag2}, where these
metal-poor dwarfs dominate the number of photometric giant candidates
at magnitudes fainter than $V$ = 18 and cannot be isolated
photometrically. (2) In deep surveys of smaller areas with low
photometric precision, most objects in the giant region of the
color-color plot are dwarfs whose photometric errors have moved them
there. Color errors in \m51\ and \mt2\ need to be smaller than 0.03
mag to avoid this problem.  An example of a survey whose photometric
errors place the giant identifications under question is
the survey for extra-tidal giants around the Carina dwarf spheroidal
of \citet{srmcarina}.  Accurate photometry and spectroscopic follow-up
of giant candidates are essential when using the Washington system to
identify the rare outer-halo giants.

\end{abstract}




\section{Introduction}

Photometric searches over wide areas for rare, distant giants in the
Galactic halo are now a reality, thanks to the advent of wide-field
CCD systems. One particularly useful photometric system is the
Washington system \citep{canterna,harriscan}. Since \citet{doug84}
added the DDO ``51'' filter to allow discrimination between late-type
dwarfs and giants and re-calibrated its \fe\ indicator, this system
has steadily gained in popularity. Its broad-band filters and strong
abundance sensitivity for late-type giants \citep{doug91,pb94} mean
that it was well-suited to studies of objects of known distance such
as open and globular cluster giants. In these cases there is little
need for luminosity discrimination.

However, when seeking rare objects such as halo giants, which are seen
through a thick veil of foreground disk stars, luminosity
discrimination is vital. The \mgb/MgH feature near 5170\AA\ is strong
in late-type dwarfs and weak in giants.  The ``51'' filter (with
$\lambda_c$ = 5130\AA\ and $\Delta\lambda$ = 154\AA) is centered on this
feature.  It works well for discriminating between late G/K dwarfs of
solar or near-solar metallicity and G/K giants.  The addition of this
filter has allowed the use of the Washington system in surveys for
late G and K giants where foreground contamination by disk dwarfs is a
problem.  This is superior to other photometric methods of identifying
giants such as DDO or Str\"omgren photometry \citep{ddo,bond80} because
only one filter has a narrow passband, and it it not necessary to
observe in the ultraviolet.  Surveys well-suited to the Washington
system include:

\begin{itemize}
\item studies of giant stars in clusters at low latitude,
 where it is now possible to
remove foreground dwarf non-members \citep[eg][]{doug97}

\item surveys for rare field giants of the outer halo (Morrison et
al.\ 2000a, Paper I hereafter; Majewski et al.\ 2000b)

\item surveys for extra-tidal
giants from Galactic dwarf  satellites \citep{srm3,srmcarina}

\end{itemize}

The last two types of survey make contributions to our understanding
of the formation and subsequent evolution of the Galaxy. 
Distant halo field giants, such as the one recently discovered by our
survey at 90 kpc \citep{edo800} provide in-situ probes of the outer
halo, and are also important probes of the Galactic
potential \citep{zaritsky}. But their identification as giants must be
solid in order to make them useful -- the spatial distribution and 
kinematics of local halo dwarfs will contribute little to our
understanding of the outer Galaxy. Indeed, misidentification will lead
to serious errors.

We began work on this paper in order to understand our low success
rate at identifying halo giants for $V$ magnitudes fainter than 18. For
brighter magnitudes our success rate, confirmed by spectroscopic
follow-up, is better than 90\%, but this falls precipitously to 24\%
for $V$ magnitudes between 18 and 19 \citep{spag2}, despite comparably
good mean photometric uncertainties for these fainter stars.

A detailed investigation into the limitations of the Washington
luminosity classification may also cast light on some surprising
claims from the surveys around dwarf satellites of the Milky Way. For
example, \citet{srmcarina} have claimed  detection of  large numbers of
extra-tidal stars in the Carina dwarf spheroidal galaxy. Such a large
extra-tidal population, if confirmed, has implications not only for
galaxy evolution but also for our understanding of the dark matter content
of the dSph satellites. However, these surveys extend to much fainter
magnitudes than previously possible, and it is important to understand
possible contamination by foreground and background objects, and the
role of photometric errors, at such faint levels.

Our essential result is that there are two major problems with the Washington luminosity
classification: first,  very metal-poor foreground subdwarfs
are indistinguishable photometrically from halo giants; second,
photometric errors can scatter disk dwarfs into the halo giant
region of the \mt2/\m51 diagnostic plot.

In Section 2 we discuss possible contaminants of halo giant candidates
in the \mt2/\m51 diagram: disk and halo dwarfs, galaxies and
QSOs. With accurate \m51 photometry, the only serious contaminants are
the very metal-poor halo dwarfs whose weak lines make them
indistinguishable photometrically from halo giants. We derive a
luminosity function specifically for such metal-poor halo dwarfs and
use it, with a simple halo model, to predict the number of subdwarfs per
\degsq\ in different directions in the Galaxy.

Although foreground disk and thick disk dwarfs are neglible
contaminants if accurate photometry is obtained, the high density of
these stars, especially at low latitudes, makes the number scattered
into the giant region of the \mt2/\m51\ diagram a sensitive function
of photometric error. In Section 3 we investigate this in detail, and
describe the double effect of large errors on giant searches: not only
are dwarfs scattered into the giant region, but genuine giants are
scattered out. 

In Section 4 we use what we have learned about giant selection to
examine our own pencil-beam survey of at high galactic latitudes
\citep{spag2}. The BTC subset (obtained using the Big Throughput Camera
on the CTIO 4m in April 1999) covers 12.7 square degrees, and has
spectroscopic follow-up of approximately half of the giant candidates
\citep{spag6}. We find that foreground subdwarfs are the principal
cause of our low detection efficiency for giants below $V=18$, and
describe our plans to optimize future searches for the faint, red
giants of the extreme outer halo so that fewer subdwarfs are
misidentified as giants.

In Section 5 we discuss the survey of \citet{srmcarina} (MOPKJG
hereafter) of approximately two square degrees centered on the Carina
dwarf spheroidal galaxy. Their study found a large number ($\sim$100)
of giant candidates outside the tidal radius of \citet{ih95}, but they
were only able to obtain follow-up spectra of three of the brightest
extra-tidal candidates. The photometric errors of the MPOKJG data are
more typical of photographic than of CCD photometry, and we discuss
the possible effect of these errors on the giant selection.

\section{Use of the 51 Filter -- Disk Dwarfs and Other Problems}

\subsection{Foreground Disk Dwarfs}

\citet{doug84} showed the sensitivity of the \m51 index to luminosity
for late G and K stars, and this was confirmed and expanded by
\citet{pb94} using synthetic spectra.  The index loses sensitivity to
luminosity blueward of \mt2 = 1.1.  It holds almost no luminosity
information blueward of \mt2 = 1.0, due to the disappearance of the
MgH feature in dwarfs at this temperature. This can be seen clearly in
Figure 7 of \citet{pb94} for $T_1-T_2<$0.4 (\mt2 $<1.0$).

The \m51 index works well for discrimination between halo giants and
foreground dwarfs of the thin and thick disk (mean [Fe/H] of --0.2
and --0.5, \citet{andy,cll}). The separation between the \fe = --0.5 dwarf
and the \fe = --1.0 giant lines ranges from 0.075 to 0.23 mags at \mt2
= 1.1 and
1.8, respectively. Only large photometric errors (explored further in Section 3
below) will cause thin or thick disk dwarfs to appear in the giant
region. We define our giant selection region as that bounded by \mt2 =
1.1 (where the index loses sensitivity) and 1.8 (where the Washington
abundance calibration stops) and \m51 = --0.02 and 0.09, 
but excluding the area below the line   
between (\mt2,\m51) = (1.1,0.02) and (1.2,--0.02). (The giant
selection region is
illustrated below in Figures \ref{badphot} and \ref{dwarflines}.

\subsection{Foreground Subdwarfs}

\begin{figure}
\includegraphics[scale=0.8]{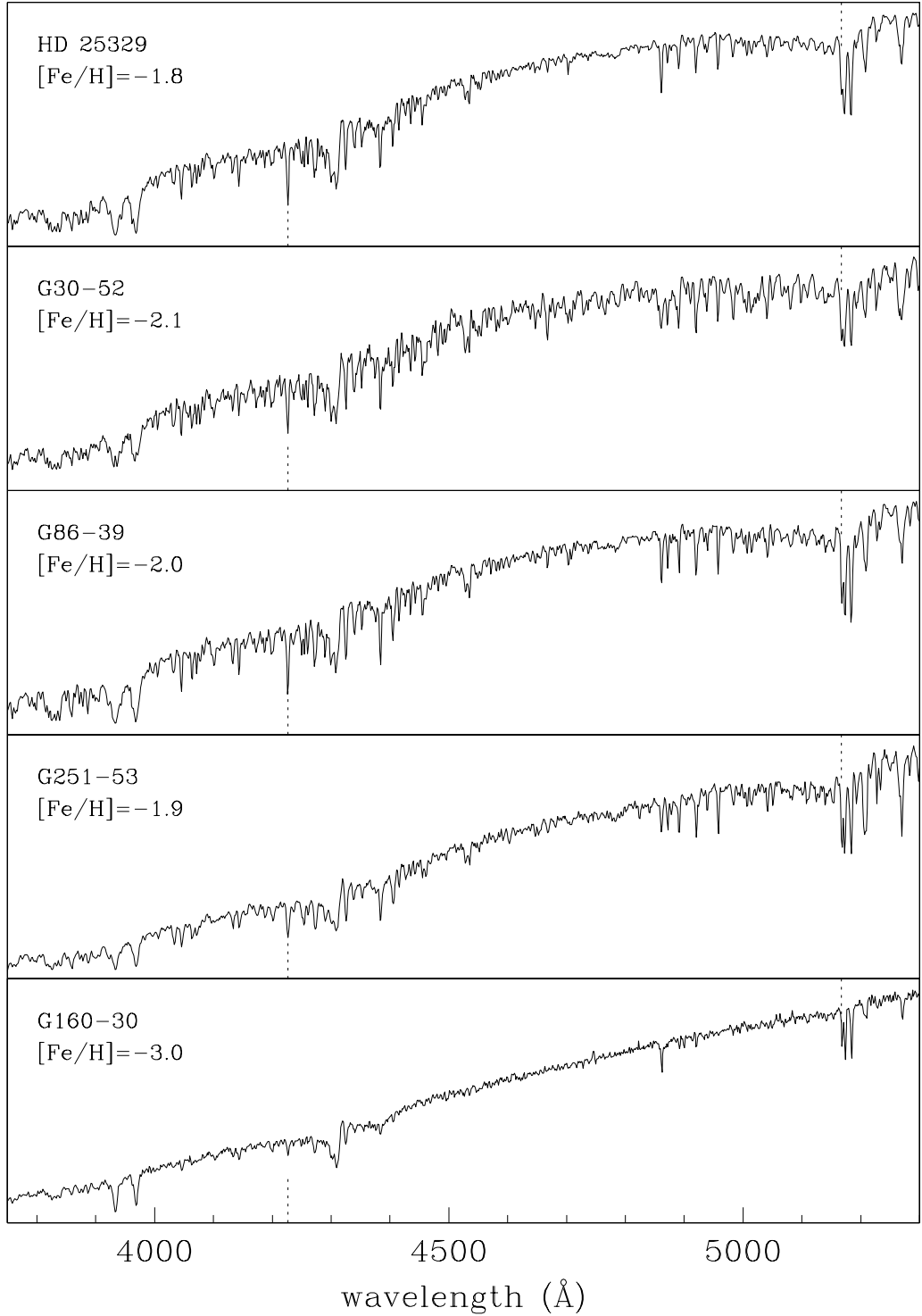}
        
\caption{Spectra of known subdwarfs in our giant color range with \fe\
less than --1.7. Positions of the CaI 4227 \AA\ line and the Mg line at
5167 \AA\ are marked.
\label{subdwarfs}}
\end{figure}

As we noted in Paper I, metal-poor dwarfs of \fe $\leq$ --2.0 
overlap the giant region in the \m51 vs. \mt2 diagram, and follow-up
spectroscopy is needed to weed them from the sample. 
Few bright examples of such stars are known, due to their intrinsic
rarity and low luminosity, and at the time of writing Paper I we had
only obtained spectra of one such star, G30-52, \fe=--2.06,
\citep{clatest}, which lies at the extreme blue end of our
selection region. Since then we have obtained spectra of several more
very metal-poor K dwarfs, which are shown in Figure
\ref{subdwarfs}. All but the most weak-lined star, G160-30, have
metallicity determinations from high-dispersion spectroscopy, whose
sources are given in Table \ref{subdwfe}.

\begin{deluxetable}{lcllccll}
\tablewidth{0pt}
\tablecaption{Metallicity and color of stars
in Fig. \ref{subdwarfs}}
\tablehead{
\colhead{Star ID} & \colhead{[Fe/H]}  &
\colhead{$(b\!-\!y)_0$} &   \colhead{(\vi)$_0$} &\colhead{$(B\!-\!V)_0$}& \colhead{(\mt2)$_0$} &\colhead{$E(B\!-\!V)$}& \colhead{Source}
}
\startdata

HD 25329 & --1.76 & 0.529 & \nodata  & \nodata & 1.10 & 0.04 & 2,3,7,8,9 \\

G30-52 & --2.1 & 0.498 & \nodata &\nodata &1.10 & 0.00 &1,2,5 \\

G86-39 & --2.0 & 0.522 & \nodata &\nodata &1.20 & 0.00 & 1,2,3 \\ 

G160-30 & --3.0 &\nodata &\nodata &1.00 & 1.50 & 0.11  & 2,5\\

G251-53 & --1.87 & \nodata &  1.24 & \nodata & 1.57 & 0.00 & 2,3,4,6 \\

\enddata
\tablecomments{
1: \citet{carbon};
2: \citet{clatest};
3: \citet{alonso};
4: \citet{sean92};
5: \citet{schlegel};
6: \citet{inese}
7: \citet{bevsned94};
8: \citet{tomlam99};
9: \citet{gratton97}
}
\label{subdwfe}
\end{deluxetable}

We have made an initial estimate\footnote{using synthetic spectra
computed by one of us (JEN)} from our low resolution spectra and find
the metallicity of G160-30 to be  \fe =
--3.0. This star is particularly useful for calibrating our survey
because of its relatively cool temperature ((\bv)$_0$=1.0, using the
reddening estimates of \citet{schlegel} and the photometry of
\citet{clatest}).  At first sight it is difficult to distinguish
between the spectra of G160-30 and that of a very metal-poor giant
such as NGC 5053-D (see Figure 3 of Paper I).  We will discuss this
further in \citet{spag6}, where we will show that the ratio of
strengths of neutral and ionized Ca lines discriminates between such giants
and dwarfs in spectra of sufficient S/N.

In Paper I we made an estimate of the number of subdwarfs with \fe\
less than --2.0 in our survey's magnitude range using the
Bahcall-Soniera model \citep{bsmodel} scaled by the proportion of halo
stars with \fe $<$ --2.0.  In the Appendix we 
derive a luminosity function (LF) specifically for the halo subdwarfs which
appear in our giant selection region.  Since
the determination of the halo field LF of \citet{bc86} only contains
51 stars over all abundances in the absolute magnitude range of
interest, we cannot derive an accurate LF from the metal-poor
subset of the \citet{bc86} data. Thus we choose to use accurate LFs
constructed from HST observations of four very metal-poor globular
clusters, and use the HST color-magnitude diagram of NGC 6397 \citet{king98}
to transform from $M_V$ to \vi\ and hence \mt2. Details are given in
the Appendix.

How distant will these contaminating subdwarfs be? Using the deep
$V$,\vi\ color-magnitude diagram of NGC 6397 of \citet{king98},
transformed to the standard Landolt system following \citet{holtz} and a
distance modulus of $(m-M)_V$=12.36, we find that \mt2=1.2 corresponds
to $M_V$=7.1. In a survey such as
ours, currently limited to $V<19$ by spectroscopic follow-up, this
corresponds to distances up to 2.5 kpc from the Sun.

\begin{figure}[h]
\includegraphics[scale=0.7]{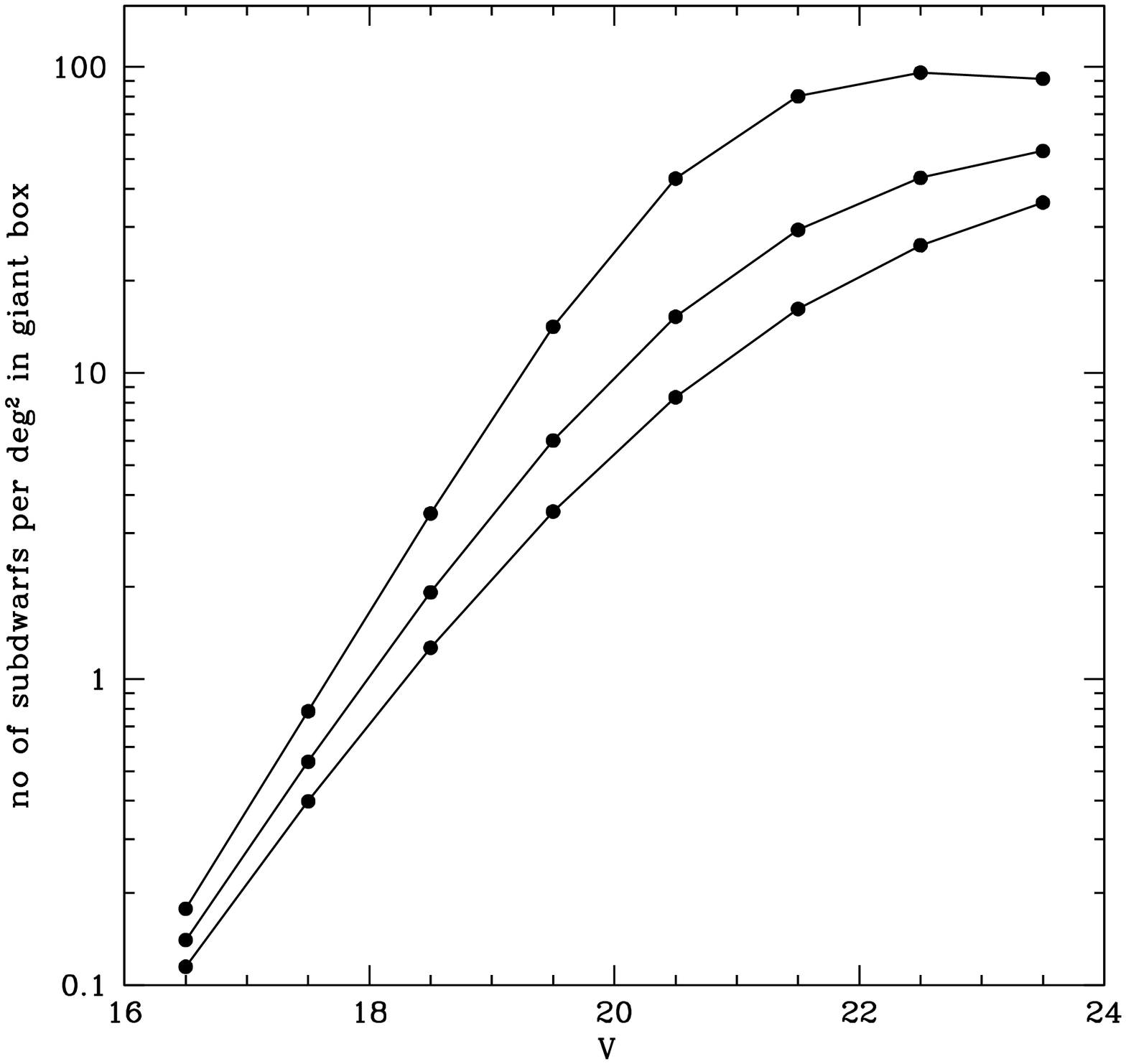}
        
\caption{Predicted numbers of subdwarfs per unit magnitude interval that
will appear in our giant region of the \mt2 vs. \m51\ diagram. Top line
is for $(l,b)$=(0,45), middle line for $(l,b)$=(90,45) and bottom line
for $(l,b)$=(180,45). There is less dependence on galactic latitude than
longitude, as the $r^{-3}$ density law of the halo contributes most
strongly to the
increase in counts in fields towards the galactic center.
\label{subdmodel}}
\end{figure}

Our halo dwarf model assumes the luminosity function given in the
Appendix in Table
\ref{lftable}. The halo's spatial distribution is modelled using
 a power-law exponent of --3.0 and a
moderate flattening of b/a=0.6 --- values confirmed by halo turnoff stars
found in our survey \citep{spag1}.
Figure \ref{subdmodel} shows its predictions for our
giant selection box. The number of subdwarfs is a strong function of
magnitude, reaching tens per \degsq\ for $V\sim20$.

\afterpage{\clearpage}

\subsection{Binary Stars}

Since up to 50\% of all stars may be binaries, it is worth considering
whether two stars in a binary system can masquerade as a single
giant. This is trivially true of a binary system consisting of a giant
and a faint dwarf companion, since the giant luminosity will
dominate. However, it is a little less clear whether a binary system
composed of two main sequence stars could have \mt2\ and \m51 colors
within the giant region. 

To test this we took the synthetic \mt2\ and \m51\ colors of
\citet{pb94} and of \citet{msb00} for stars with $T_{eff}$ ranging
from 4000 to 10000 K, $log g$ from 4.5 to 0.7, and \fe\ from 0 to
--1.0, and combined them into simulated binary systems using the
absolute magnitudes given in \citet{msb00}. Since \citet{pb94} only
give colors for stars from 4000 to 7000 K, and \citet{msb00} does not
give \m51\ colors, we set the \m51\ color of all stars hotter than
7000K at 0.017, which is the color expected for a spectrum with no
features in the \mgb/MgH region. The only stars for which the binary
system had both \mt2\ {\it and} \m51\ in our giant region were those
for which one component of the binary system already had colors within
the giant region. This is because (1) the bluer component of a binary
made of two main sequence stars will be brighter and so dominate the
\mt2\ color, and (2) it is not possible in a spectrum without emission
lines to have \m51\ significantly higher than 0.017 unless one star is
extremely metal-poor.

Thus contamination by binary stars whose constituent stars are both on
the main sequence is negligible.

\subsection{Galaxies and QSOs}

Because of the emission lines and redshifted features in galaxy and
QSO spectra, they can be found both in the giant region of the
\m51/\mt2\ plot, and also in regions not occupied by normal stars.
The synthetic \m51 colors of \citet{pb94} for very metal-poor stars
reach no higher than \m51=0.05,  corresponding to the
value that a blackbody spectrum would have.  Thus, apart from
photometric errors, we would not expect to see any normal or
metal-poor stars with \m51 $>$ 0.05. The only other possibility is a
spectrum which has {\it less} light, on average, in the wider M band
than in the narrow 51 band close to its center. This can happen if
absorption lines in a galaxy's spectrum are redshifted into the M
band, or if an emission line in a galaxy or QSO spectrum is redshifted
into the 51 band. 

\begin{figure}[h]
\includegraphics[scale=0.8]{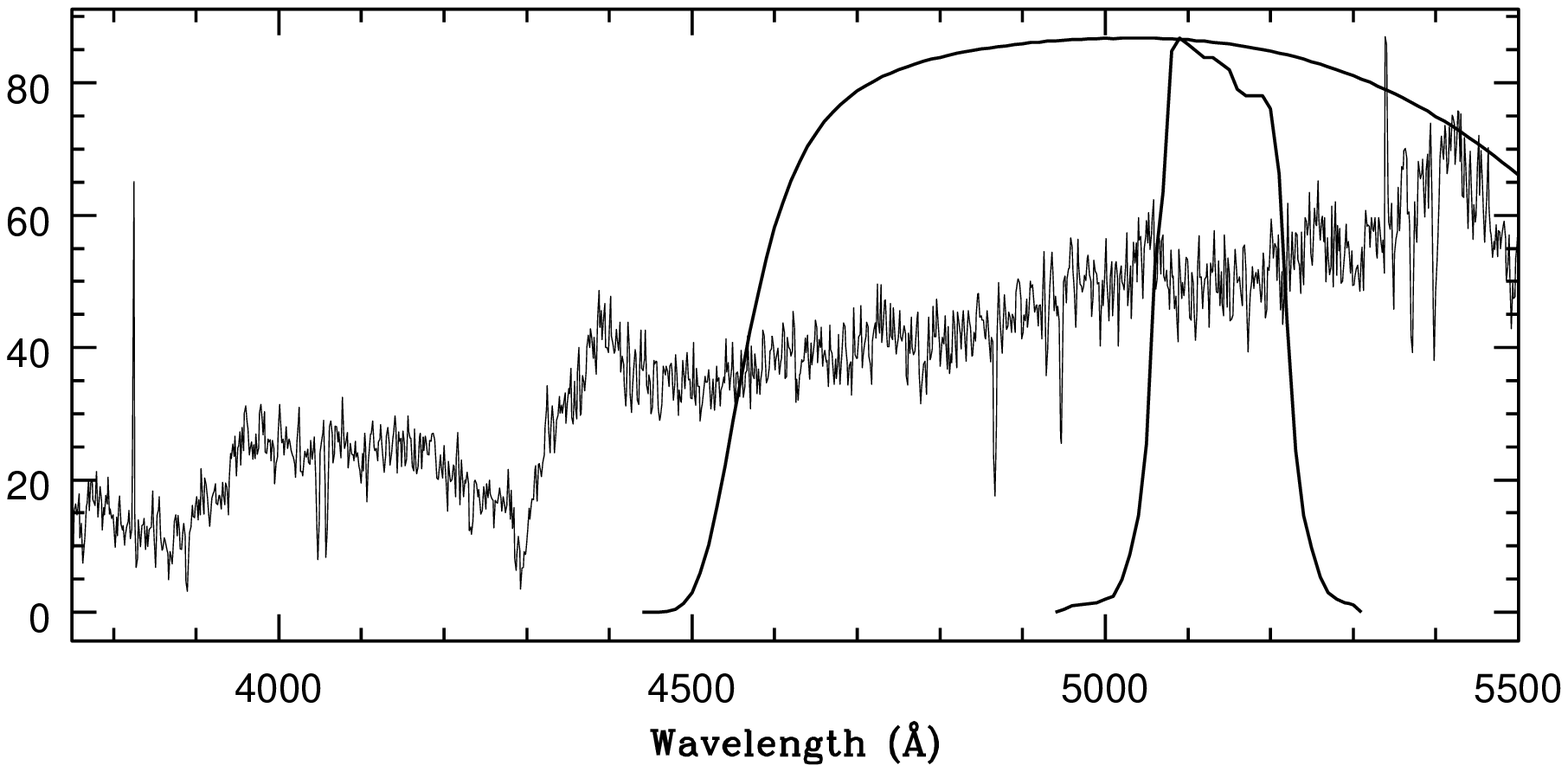}
\caption{Spectrum of BAL QSO at RA=9:16:16.97 Dec=12:16:09.6 (2000),
with the $M$ and narrower 51 passbands overlaid. The QSO's relatively
featureless spectrum across the $M$ passband, and the broad emission
line at the blue end of the 51 band, contribute to its position in the
giant box of the \mt2/\m51\ diagram.
\label{qso}}
\end{figure}

Figure \ref{qso} shows the spectrum of a BAL QSO identified in our
survey, with the $M$ and 51 filter passbands overlaid\footnote{the 51
filter passband should be shifted a few \AA\ to the blue  to
compensate for the fast beam of the CTIO 4m prime focus.}. The QSO
spectrum is, to first order, featureless over the $M$ passband, although
part of the broad emission line at the blue end of the 51 passband
would have been
included in this filter's observations, giving the same or slightly
more light per \AA\ in the 51 filter than the wider $M$ band.

In addition, the Washington colors of some normal galaxies are similar
to those of metal-poor stars, so we will find a small number of
galaxies in other regions of the \mt2 vs. \m51 diagram. Our survey
uses data from mosaic CCD cameras (the Big Throughput Camera and the
NOAO mosaics)  on the KPNO and CTIO 4m telescopes
and the Burrell Schmidt. We are
particularly vulnerable to contamination by galaxies in our Schmidt
data, where the large pixels make it more difficult to distinguish
between stars and galaxies. There is still a small chance of
contamination by compact galaxies and QSOs even in our better-sampled
BTC and mosaic data. We discuss this further in Section 4.

\section{Photometric Errors}

Use of the \m51 index for luminosity classification requires accurate
colors for two reasons. First, the separation between giants and
dwarfs in the \m51 vs. \mt2 plot is not large: $\sim$ 0.15 mag between
solar abundance dwarfs and giants for \mt2 $>$ 1.3 (\bv $>$ 1.0) and
less for bluer stars. Although metal-poor giants lie further away from
the dwarf locus because of their weak \mgb\ lines, this advantage is
offset by the fact that as we move to fainter magnitudes, foreground
dwarfs become dominated by thick disk stars, whose lower metallicity 
([Fe/H]$\sim-0.5$, \citet{cll}) moves them toward the giant region. Second, 
distant giants are strongly outnumbered by foreground disk stars, so
we need to cope with the effect of 2$\sigma$ and even 3$\sigma$ errors on the
dwarf colors.

We use the accurate Washington photometry obtained with the Big
Throughput Camera (BTC) on the CTIO 4m telescope in April 1999
\citep{spag2} to illustrate the effect of photometric errors on
giant/dwarf classification. This pencil-beam data covers 12.7 square
degrees, with latitude ranging from $b$ = 25 to 72. For $V<19$, the
median error in \mt2 was 0.012 mag, and for \m51, 0.013 mag.

\begin{figure}[h]
\plotone{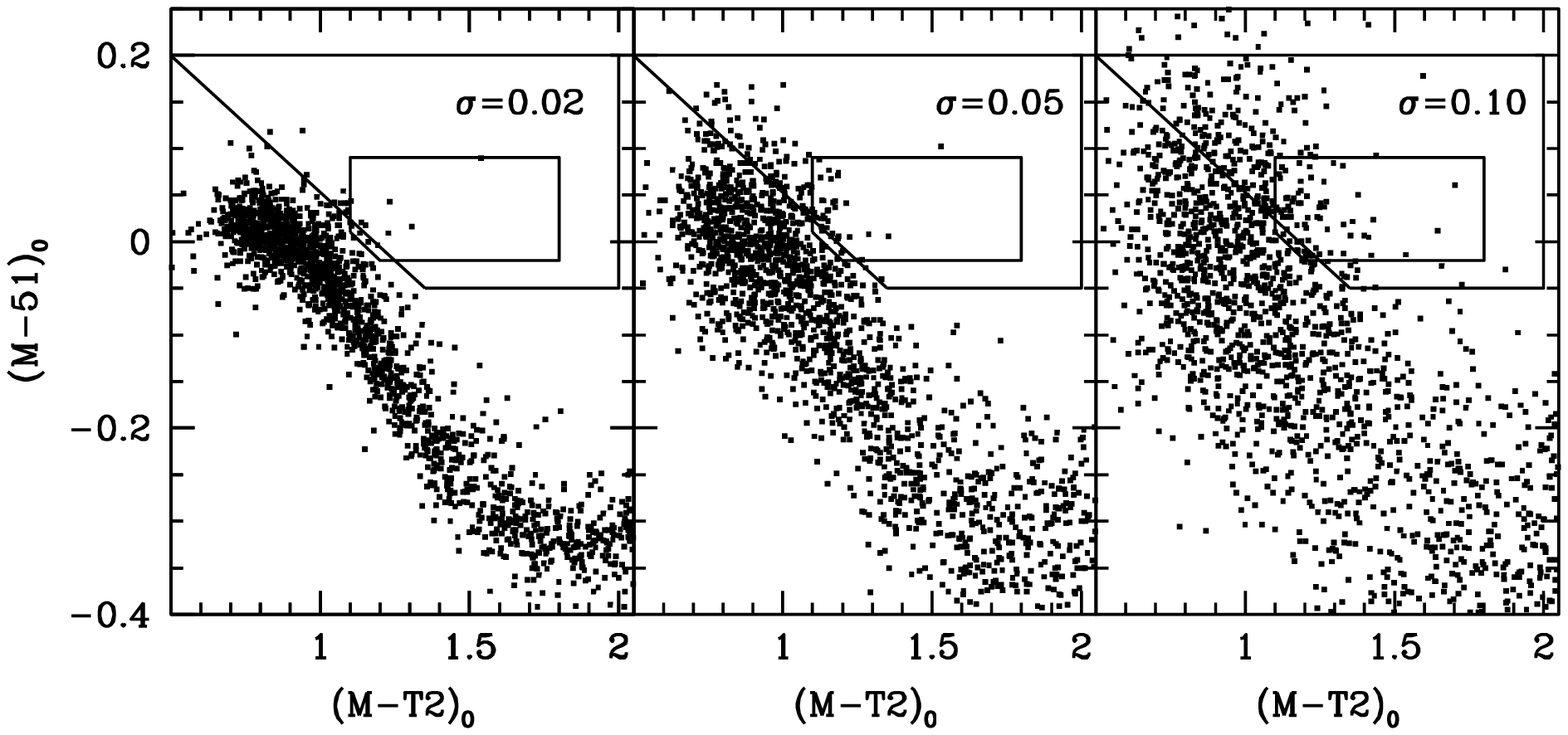}
        
\caption{\mt2 vs \m51 diagrams for BTC data degraded with increasing
color errors from left to right. Left-hand panel shows the original
data, which has a clearly delineated region occupied mostly by
foreground disk stars, middle panel shows the same data with 5\% color
errors in both \mt2 and \m51, right-hand panel show the data with 10\%
color errors in both filters. The boxes used for giant classification
by our collaboration and by MOKP are marked -- the MOKP box is larger
and stretches significantly bluer.
\label{badphot}}
\end{figure}

We chose a subset of these data with low errors using a magnitude
range of $V=17-18.5$ and a maximum error of 0.02 mag in both \mt2 and
\m51. This subset will be used as an approximation of ``zero-error''
data which will then be degraded with successively larger photometric
errors by adding a number chosen from a gaussian error distribution.
 The left-hand panel of Figure \ref{badphot} shows these data, with
the regions used by our collaboration and \citet{srm1}(MOKP
hereafter) for giant selection marked. Our giant region is significantly
smaller and does not extend as far blueward as the MOKP region. It can
be seen that the most numerous stars in this  color range have \mt2
between 0.7 and 1.2 (\bv $\sim$ 0.5--0.9). These stars are mostly disk
dwarfs \citep{bsmodel}. 

We simulated photometric errors by adding gaussian\footnote{Often in 
astronomical datasets we do not have detailed information on
the error distribution shape. In general, an
error distribution with larger wings will produce more bogus giants,
and a distribution with smaller wings than a gaussian will have less
effect.}  errors to the
observed colors in this low-error subset. The color error in \m51 and in \mt2 
was chosen randomly from a
gaussian distribution with either $\sigma$ = 0.05 (middle panel) and 0.10
(right-hand panel) mags. 
It can be seen that even
0.05 mag color errors move stars into both ``giant'' boxes, in
particular into the region of the MOKP box with \mt2 $<$ 1.0.  Color
errors of 0.10 mag produce large numbers of bogus giants.

\subsection{Where Do the Stars Go?}

To quantify the effect of photometric errors on our giant
classifications, we need to explore two effects: genuine giants
removed from the giant region, and dwarfs and other stars
scattered into the giant region by color errors.

\begin{figure}
\includegraphics[scale=0.6]{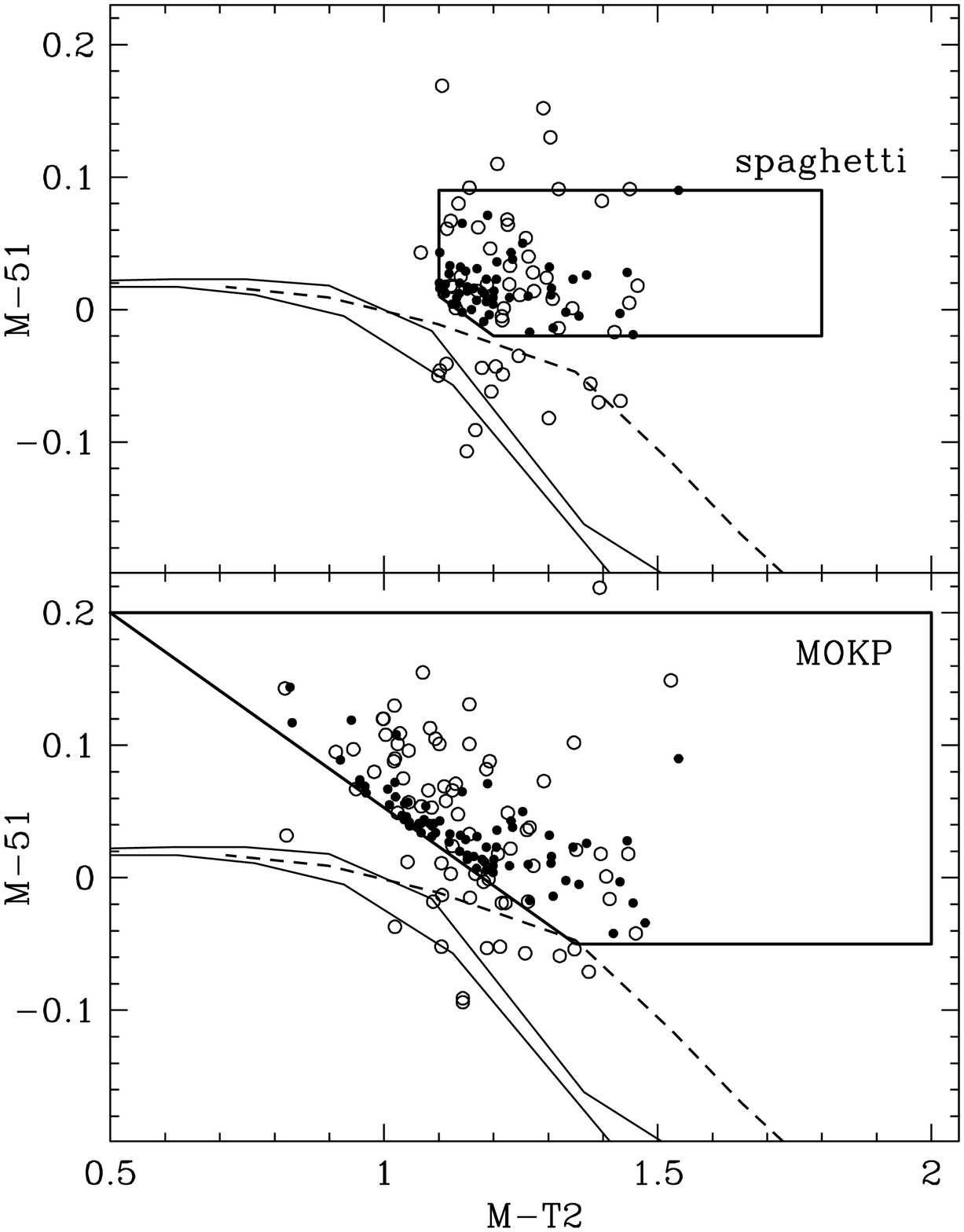}
        
\caption{For each classification scheme, the movement of stars
originally classified as giants in that scheme is shown. Solid circles
are original photometry, open circles show these data degraded with a 0.05
mag gaussian error. The box in the top panel shows our (``spaghetti'') classification
scheme, with solid lines showing dwarf loci from \citet{pb94} with
[Fe/H] = 0.0 and --1.0, and dashed line the solar abundance giant
locus from the same paper. The box in the bottom panel shows the
classification scheme of MOKP, with the same symbols. It can be seen
that more stars are lost out of our giant box because it is smaller.
\label{losegiants}}
\end{figure}
\afterpage{\clearpage}

\begin{figure}
\includegraphics[scale=0.6,angle=270]{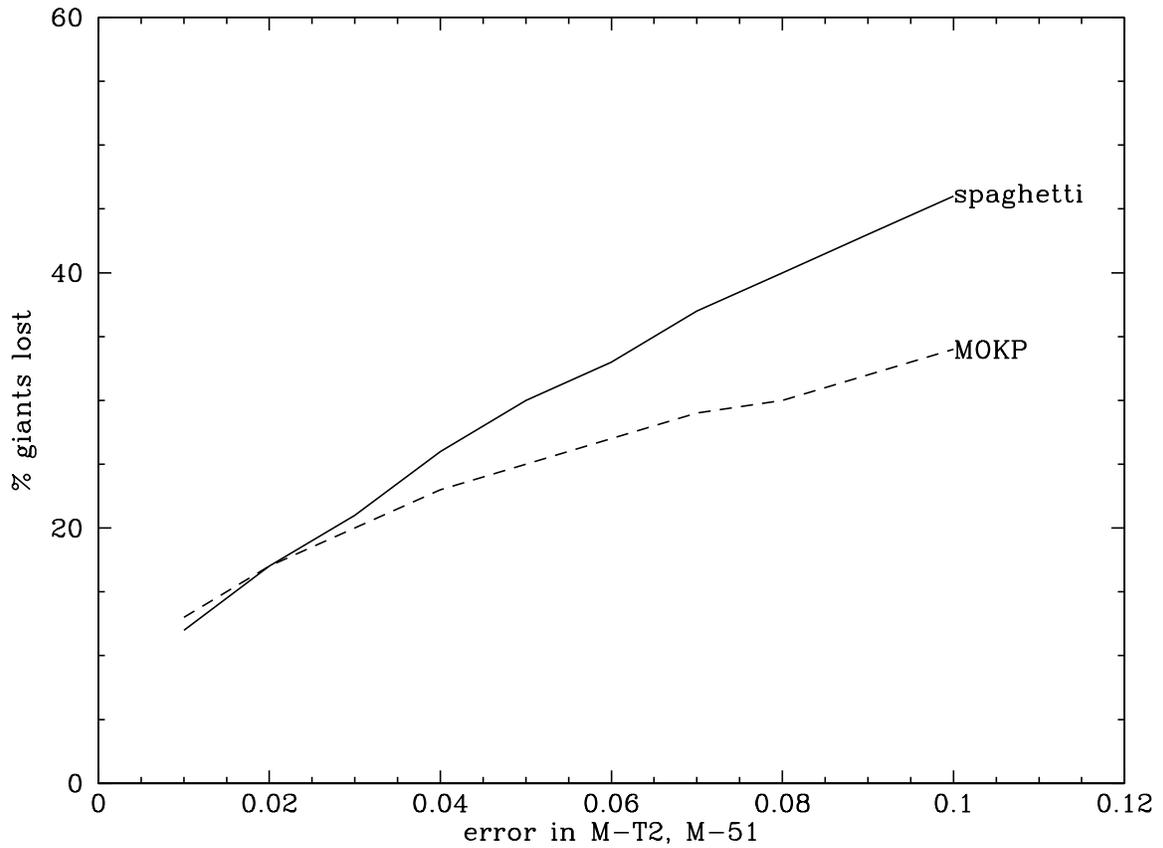}
        
\caption{Percentage of genuine giants lost due to photometric
errors, for our (``spaghetti'') and the MOKP classification schemes. The MOKP scheme loses less
giants because of its larger giant classification box.
\label{losegiantsumm}}
\end{figure}

The first effect is simpler to quantify because it does not depend on
the position in the Galaxy of the field surveyed. Figure
\ref{losegiants} illustrates the movement of genuine giants in the
\m51/\mt2\ diagram for the two classification schemes, assuming that
each color was degraded by an amount drawn from a Gaussian with
$\sigma$ = 0.05 mag.
Our smaller giant region leads to a higher
loss rate of giants from the box.  Figure \ref{losegiantsumm}
summarizes the percentage of giants lost due to errors, as a function
of the color error. It can be seen that even for color errors as low
as 0.02 mag, around 20\% of genuine giants will be lost. The most
numerous (and least luminous) giant stars lie close to the blue
boundary of our giant box at \mt2 = 1.1, and the majority of stars will
be lost from this region.  This effect, and its dependence on color
and absolute magnitude, will need to be taken into account when using
halo giants found with \m51 to measure the density distribution of the
outer halo.

\begin{figure}
\includegraphics[scale=0.7]{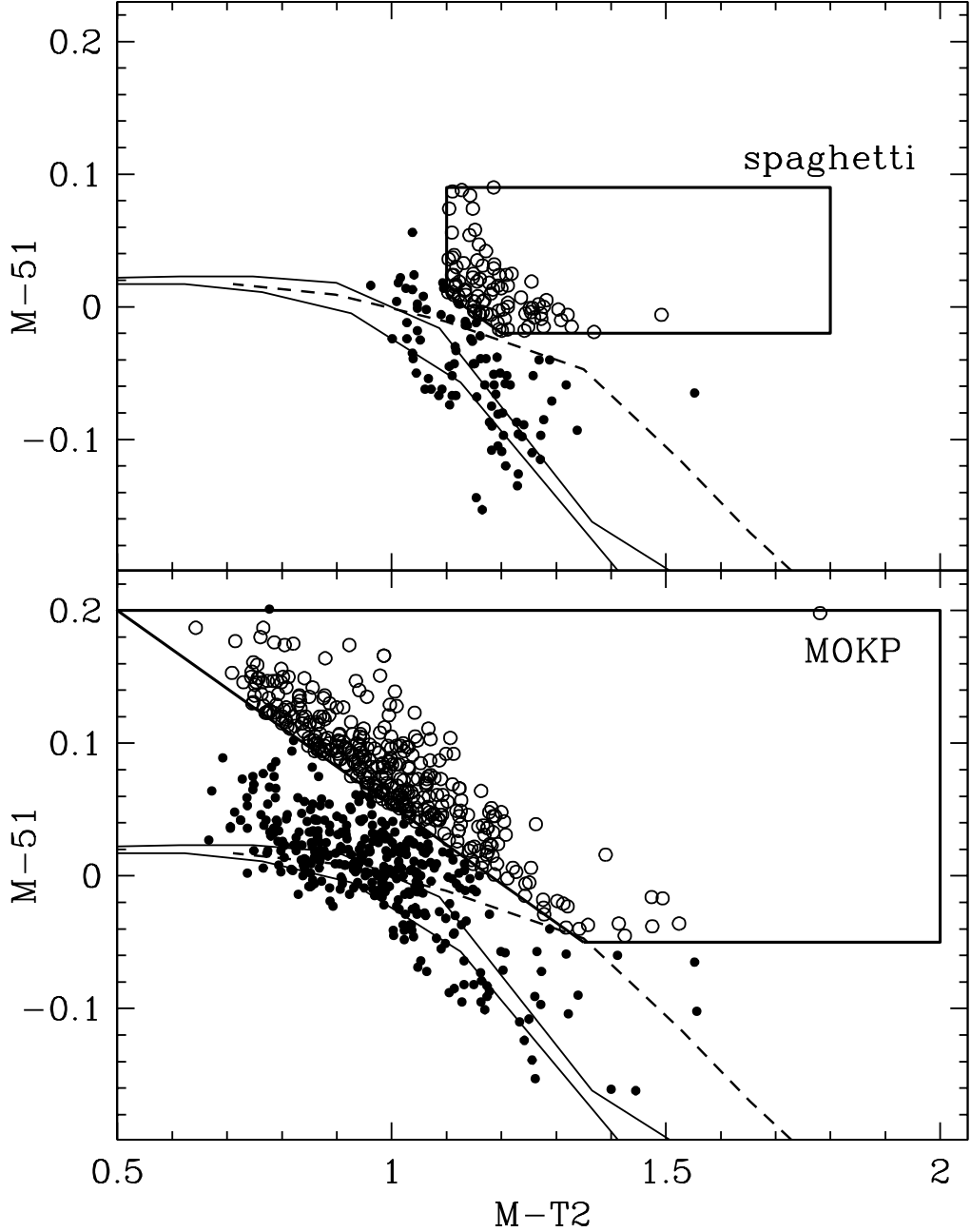}
        
\caption{Regions where bogus giants, wrongly classified because of
their photometric errors, originate for (top panel) our classification
scheme and (bottom panel) that of MOKP. Solid symbols denote original
colors, open symbols the corresponding point when colors are degraded
with a 0.05 mag error drawn from a gaussian distribution. The giant
classification box is outlined by a thick solid line in each
panel. Thin solid
lines show the solar abundance and [Fe/H]=--1.0 dwarf loci from the
synthetic spectra of \citet{pb94} and the dashed line the solar
abundance giant locus from the same work.
\label{dwarfsin}}
\end{figure}

The leakage of foreground dwarfs into the giant regions is more
complex to quantify, as it depends on the numbers of dwarfs close to
the boundaries of the giant box. This is a strong function of
galactic latitude in particular, since the path length of lower-latitude
lines of sight through the disk is higher. Figure \ref{dwarfsin}
illustrates the regions of the color-color diagram where bogus giants
originate for both classification schemes, in the case of 0.05 mag
errors. (Only stars that will be observed in the fiant selection
regions will be followed.)
For our classification scheme, most of the interlopers are dwarf stars
with \mt2 = 1.0--1.3 (\bv = 0.7--1.0). For the MOKP
scheme, many more stars are scattered into the giant region. In
particular, the blue extension for \m51 $>0$ causes problems.  The
vast majority of stars that find their way into the giant region have
\mt2 $<$ 1.0.  For stars this blue, the \m51 index has little or no
sensitivity to luminosity, as the \mgb/MgH lines become weak for
dwarfs and giants alike. The convergence of the dwarf and giant
sequences can clearly be seen in Figure 7 of \citet{pb94} for
$T_1-T_2<$ 0.4 (\mt2 $<1.0$), and Geisler's original (1984) paper does
not even show data for $T_1-T_2<$ 0.38.

\begin{figure}
\includegraphics[scale=0.65,angle=270]{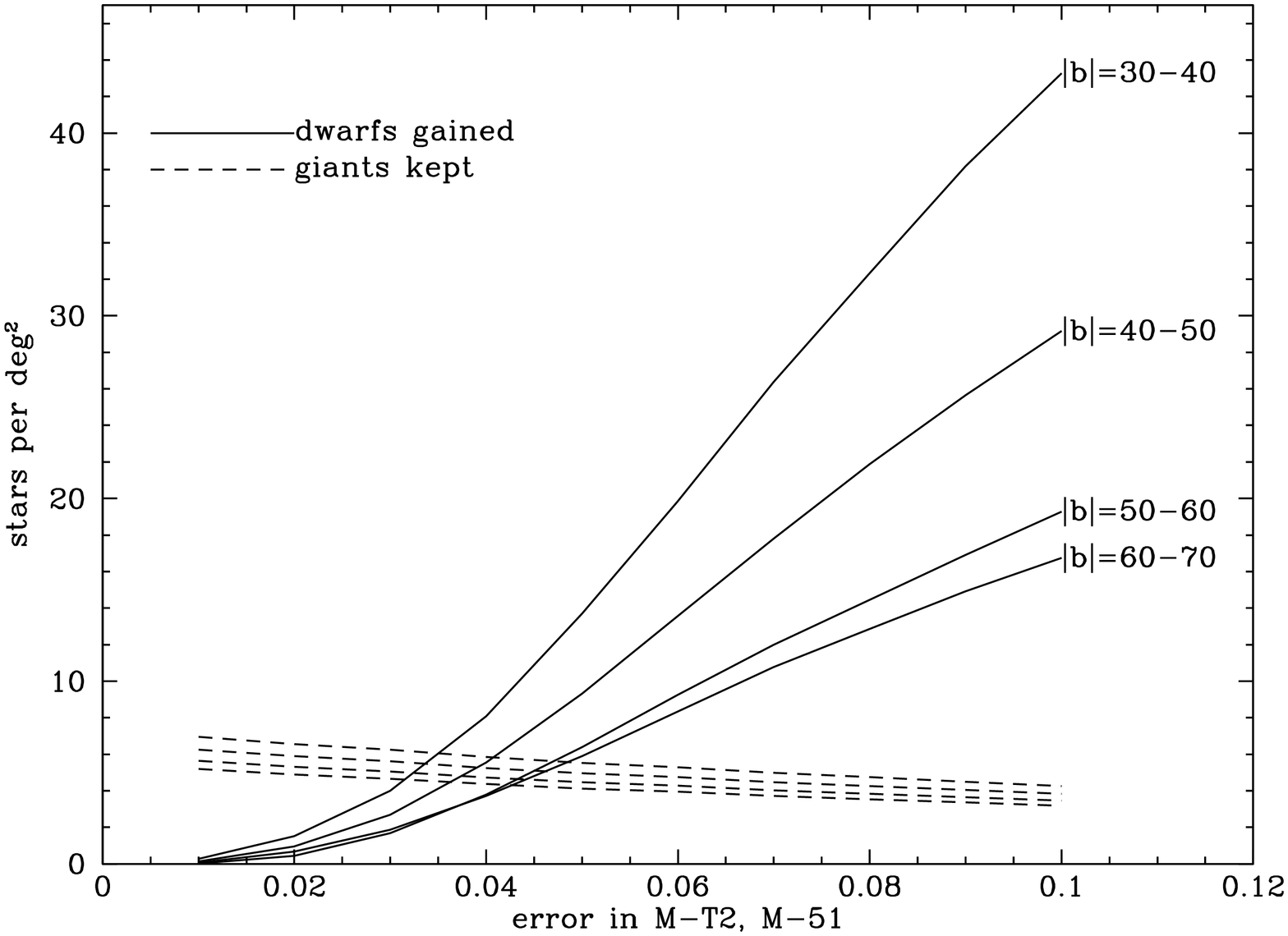}
        
\caption{The tradeoff between losing giants and
gaining dwarfs via photometric errors, for our giant classification
region. The BTC data for $16<V<19$ were divided into ranges of latitude and both
giant and dwarf loss/gain lines were plotted separately for each
latitude bin. The giant lines are not labelled for space reasons, but
the lowest latitude is denoted by the upper dashed line and the
highest latitude by the  lowest of the four dashed lines.
\label{ourbal}}
\end{figure}

\begin{figure}
\includegraphics[scale=0.65,angle=270]{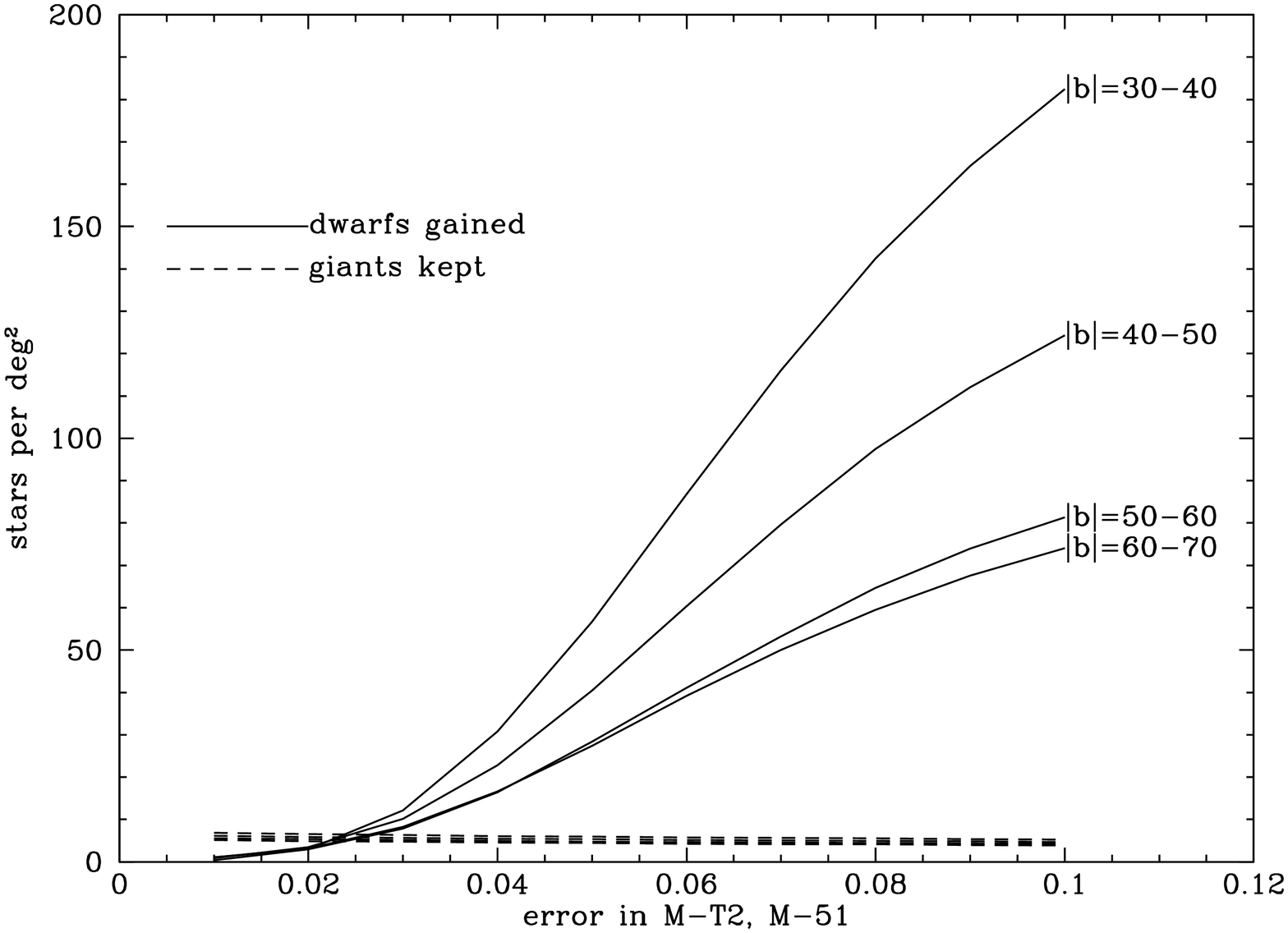}
        
\caption{Same as Fig. \ref{ourbal} but for the MOKP giant selection
scheme. Note the change in scale from Fig. \ref{ourbal}.
\label{srmbal}}
\end{figure}

Figures \ref{ourbal} and \ref{srmbal} summarize the balance between the
number of giants remaining in the giant region and the number of
dwarfs moved into the giant region by photometric error, as a function
of color error and galactic latitude.  For these simulations we used
the BTC data without error cuts, in the magnitude range V=16--19, which
is typically the magnitude range we use to select giants in our survey
(as spectroscopic follow-up becomes difficult fainter than V=19 on
4m-class telescopes). The absence of error cuts is because we need to
measure how many dwarf stars are in the correct color and magnitude
ranges to be moved into the giant region via photometric errors, and
limiting the data selected via our errors will lead to incompleteness
in our sample. (Note that at these
magnitudes the BTC data are essentially 100\% complete.)
We define a star as a dwarf if it is below the solar abundance giant
line in the \mt2 vs. \m51 diagram or if it is bluer than
\mt2=1.0. This latter requirement will include some blue giants, but
the \m51 color holds no information about luminosity this blue.
In order to give a feeling for how important this is with respect to
the actual numbers of halo giants likely to be found (in a smooth
halo) we have used the predictions of the halo model of \citet{hlm93}
for each latitude bin, and the same magnitude range. (These predictions
agree in general with the numbers of halo giants that we have found in
our survey fields.) The number of halo giants falls as the error
increases because more giants get scattered out of the halo region. 
	
It can be seen from the plots that, especially for low latitudes,
large numbers of dwarfs are scattered into the giant box. For our
classification scheme, the 50\% success point where the number of dwarfs scattered in equals the
number of giants not scattered out of the giant box occurs 
for color errors between 0.035 and 0.045
mag. For the larger giant box of MOKP, the effect is extreme -- bogus
giants outnumber genuine ones for color errors greater than 0.02
mag The number of giant candidates that are actually dwarfs can reach
more than 100 per square degree for large errors and lower latitudes.
It is clear that spectroscopic follow-up of every giant candidate to
confirm its luminosity is vitally important. Without it, spurious
conclusions will surely result.

With this background behind us, we now consider implications for our
own survey and that of MOPKJG.

\section{Example 1. Spectroscopic Followup of our BTC Data}

Here we describe our spectroscopic follow-up of the BTC imaging
data of 12.7 square degrees \citep{spag2}, focussing on the dwarfs and
other contaminants that we identified. In a future paper \citep{spag6}
we will describe our spectroscopic calibrations to measure luminosity
and metallicity in detail. In three observing runs on the KPNO 4m
telescope (in May 1999, January 2000 and March 2000) we observed and
classified 57 halo giant candidates, with V magnitudes ranging from
$V_0=15.6 - 19.2$. This constitutes approximately 50\% of our giant
candidates in the BTC data.   

One candidate (with $V_0$=18.7) was a QSO; so far we have found no 
galaxies in our BTC data, due to the good spatial sampling of the BTC CCDs.

\subsection{Foreground Dwarfs}

We have estimated the metallicity of the foreground dwarfs that we
identified, by comparing their \mgb/MgH line strength at the observed \mt2
color with that of the dwarf and subdwarf standards observed on these
runs (see Table \ref{subdwarfs} above and Table 2 of \citet{spag1}).
This allowed us to classify them into subdwarfs with \fe $<$ --2.0 or
metal-richer dwarfs and subdwarfs. Figure 8 of \citet{spag2} shows
their position in the \mt2 vs. \m51 diagram. Most of the foreground
stars that we identified were metal-poor subdwarfs, confirming the
efficacy of our luminosity discrimination using accurate \mt2 and \m51
colors. Five stars out of the 57 were classified as dwarfs with \fe\
greater than --2.0: two of these were observed before we fully
understood the strong effect of errors in scattering dwarfs into our
giant box. Since these two stars have errors in both \mt2 and \m51 of 0.04--0.05 mag,
it is not unexpected in the light of  Section 3
above that they are foreground dwarfs. The other three dwarf stars
have low errors and $V_0 >$ 18, and this is consistent with the results of Figure
\ref{ourbal}: even with errors of 0.02 mag in both colors (a
little higher than our mean errors at this magnitude), approximately
one dwarf per square degree will be scattered into our giant
region. (We have observed approximately six square degrees
spectroscopically in this field.)

\subsection{Extreme Subdwarfs}

\begin{figure}
\includegraphics[scale=0.5,angle=270]{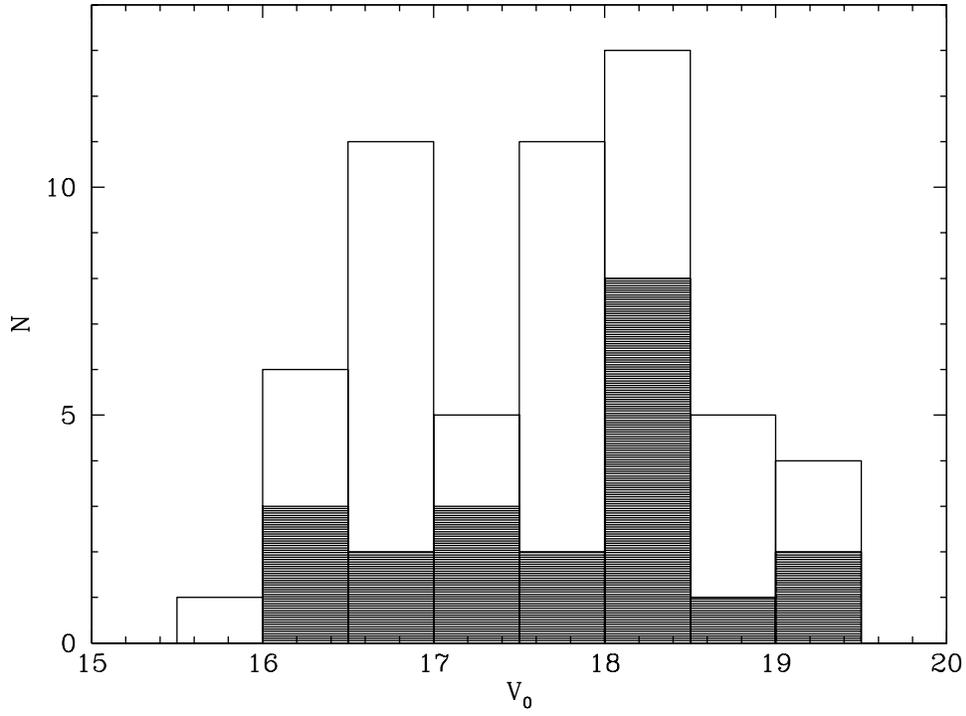}
        
\caption{Magnitude distribution of the halo giant candidates for which
we obtained follow-up spectra, with the histogram of magnitudes of the
subdwarf stars shown shaded.
\label{subdwhist}}
\end{figure}

Out of the 57 halo giant candidates observed, 22 were subdwarfs. 
(Our giant selection criteria have evolved
as we learned more about metal-poor subdwarfs, so some of these
stars have colors which place them outside our giant selection
box.) Figure \ref{subdwhist} shows the histogram of V magnitudes of
the 57 halo candidates observed, with the histogram of subdwarf
magnitudes overlaid.
The magnitude distribution of halo candidates observed reflects the
weather and seeing during our 4m spectroscopic runs, rather than the true
distribution of halo giants. The sample size is small, but it is still
possible to see that the proportion of subdwarfs increases as we go to
fainter magnitudes.

\begin{figure}
\includegraphics[scale=0.7]{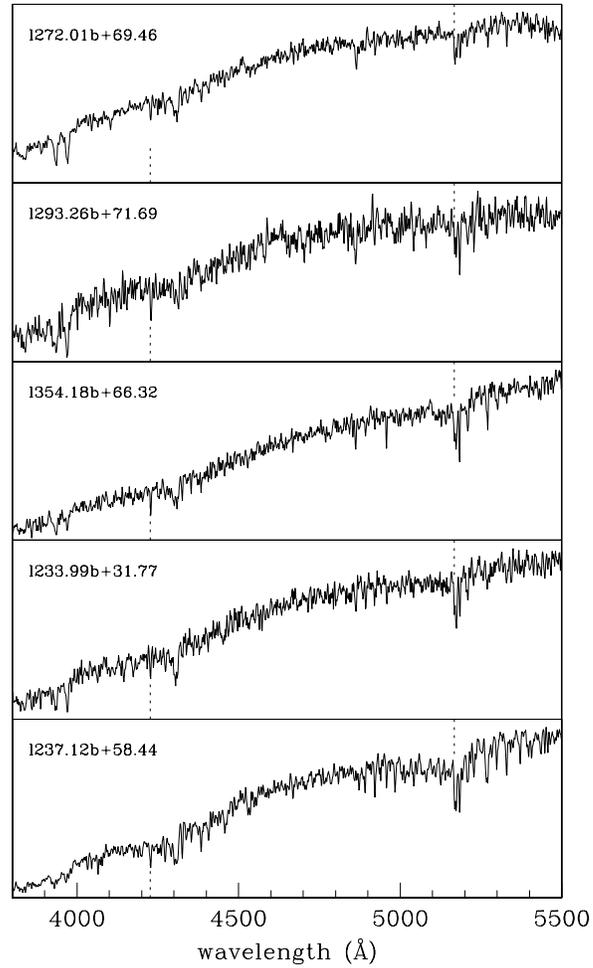}
        
\caption{Spectra of subdwarf stars identified in our halo survey.
Positions of the
Ca I 4227\AA\ line and the Mg 5167\AA\ line are marked. 
\label{oursubdwarfs}}
\end{figure}

Figure \ref{oursubdwarfs} shows some examples of spectra of subdwarf
stars identified in our BTC survey. Photometry and coordinates for these
stars is given in Table \ref{oursubdw}.  Spectral features similar
to those of the subdwarf standards shown in Figure \ref{subdwarfs} are
clearly seen: a weak  MgH feature and strong CaI 4227
line. 

\begin{deluxetable}{lrrlccccc}
\tabletypesize{\scriptsize}
\tablewidth{0pt}
\tablecaption{Photometry for subdwarf stars in Fig. \ref{oursubdwarfs}}
\tablehead{
\colhead{Star ID} & \colhead{RA} & \colhead{Dec} & \colhead{epoch} &
\colhead{$M_0$} & \colhead{$(M\!-\!T_2)_0$} &
\colhead{\mt2 error}& \colhead{$(M\!-\!51)_0$} & \colhead{\m51 error}
}

\startdata
l272.01b+69.46& 12:09:20.85 &  9:09:08.6 & 2000 & 17.625 & 1.383 & 0.009 & --0.093 & 0.009 \\

l293.26b+71.69 & 12:39:11.00 & 9:03:07.5 &2000& 18.718 & 1.356 & 0.011 & --0.004 & 0.010 \\

l354.18b+66.32 &14:05:55.81 & 11:10:29.1 & 2000& 18.702 & 1.420 & 0.009& --0.042 & 0.016 \\

l233.99b+31.77 & 9:21:16.95& --1:49:28.8 & 2000& 19.183 &1.140&0.010 &  0.004 & 0.011\\

l237.12b+58.44 & 10:54:50.52& 11:24:06.4 & 2000& 18.225& 1.148&0.011&   0.018& 0.011 \\ 

\enddata

\label{oursubdw}
\end{deluxetable}
Star l354.18b+66.32 in the middle panel of Figure \ref{oursubdwarfs}
is a particularly interesting one. 
Its color and spectrum are similar to that of the subdwarf standard 
G251-53, with \fe =
--1.9. The synthetic colors of \citet{pb94} predict that this star
should have an \m51 color  several hundredths of a magnitude
lower than the bottom of our giant selection box, and its \m51 value
(--0.041) confirms this prediction. This adds confidence that the estimate of
the luminosity function for halo subdwarfs found in our giant box in
the Appendix, based on the synthetic colors of \citet{pb94}, should be
reasonably accurate.

We can make a more direct test of the accuracy of our subdwarf luminosity
function by comparing its predictions of the numbers of subdwarfs we
should find in our BTC data with the actual number found. Now we
restrict consideration to halo giant candidates {\it inside} our giant
selection box with photometric \fe\ estimate less than --1.0, and limit
the magnitude range to $V_0<19.0$.
Of the 53 BTC fields, 14 have complete spectroscopic follow-up of giant
candidates, and 8 subdwarfs were identified. Another 13 fields have no
giant candidates at all. Assigning the correct percentage of ``no
giant candidate'' fields to the spectroscopic follow-up group, we find
that 4.45 square degrees have been completely surveyed for
giants. This gives a subdwarf frequency of 1.8$\pm$0.6 per square
degree. The fields with complete spectroscopic follow-up have a mean
(l,b) = (281,47), so the middle line in Figure \ref{subdmodel} is a
good match. This predicts 2.6 subdwarfs per square degree in our giant
selection box down to $V$=19, in good agreement with our actual
value, once errors in the local halo density are taken into account. 

In summary, our spectroscopic follow-up has confirmed the importance of
accurate photometry for giant selection using the Washington
system. In addition, the number of subdwarfs found in our fields is in
agreement with the predictions of the model described in Section 2.2.1
and the Appendix. Particularly for $V$ magnitudes fainter than 18,
{\it photometric identification of halo giants is not enough}:
spectroscopic confirmation of their luminosity is very important.

\subsection{Efficiently Selecting Faint Giants}

If we observe every giant candidate that appears in the giant region
of Figure \ref{dwarflines} (see Appendix) spectroscopically for V
fainter than 18, we will find three subdwarfs for every genuine halo
giant.  The stars that are of particular interest in this magnitude
range are the high-luminosity giants which help trace the extreme
outer halo, like our star l304.49b+60.51 \citep{edo800}. With accurate
photometry, it is possible to make our spectroscopic follow-up more
efficient by utilizing the sensitivity of the \m51 color to
luminosity.

\begin{figure}
\includegraphics[scale=0.5,angle=270]{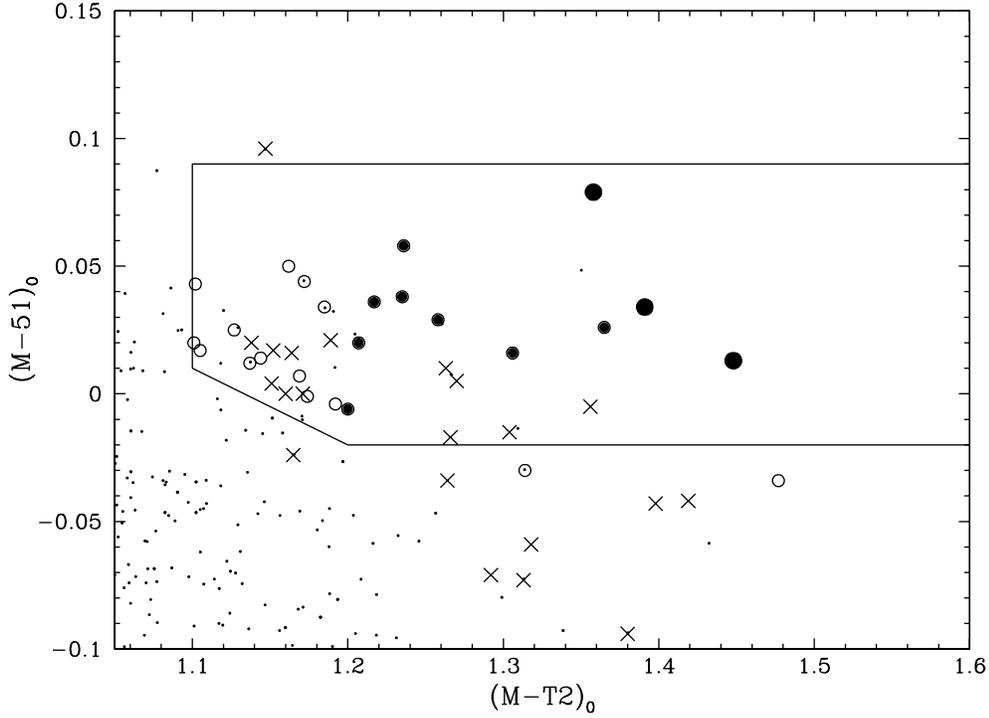}
        
\caption{Position of our most luminous giants in the \m51 vs. \mt2
diagram. Our giant selection box is outlined, and the small dots
represent the photometry from a subsample of our survey fields to show
the locus of foreground stars. Subdwarfs are shown as crosses and
giants as circles. Open circles are the least luminous giants, with
$M_V>0.5$, small closed circles are giants with $-0.5<M_V<0.5$, large
closed circles are giants with $M_V<-0.5$. The most luminous giants
are in the upper right region of our giant box while the subdwarfs
occupy the bottom left for this magnitude range.
\label{redgiants}}
\end{figure}

Figure \ref{redgiants} shows the halo giants and subdwarfs identified
by spectroscopic follow-up, with giants coded by absolute magnitude. It
can be seen that the most luminous giants are not only the reddest,
but also occupy the upper region of the giant box, with \m51 $>$ 0.01,
while the subdwarfs cluster toward the bottom left. 

For our survey, if we restrict our giant selection box to the upper
right of the original region for $V>18$, we are able to optimize our
efficiency of giant selection without missing any of the most distant
giants in our sample.  
For data that reach fainter levels, this "new, clean" region
becomes contaminated with intrinsically fainter, redder subdwarfs.

\section{Example 2. The MOPKJG Survey of the Carina Dwarf Spheroidal Galaxy}

\citet{srmcarina} present Washington M, T2, 51 photometry of a field
of area $\sim$2 deg$^2$ centered on the Carina dwarf spheroidal
galaxy, and use both the \mt2 vs. \m51 diagram and the locus of the
Carina red giant branch in the $M_0$ vs. \mt2\ diagram to identify a
large number ($\sim 100$) of Carina-associated giant candidates {\it
outside} the tidal radius of \citet{ih95}. These stars range in
magnitude from $M_0$=17.8--20.8 ($V\simeq$17.5--20.5).
They obtained confirmatory
spectroscopy of three of the brightest extratidal giant candidates,
which are  1.4$\sigma$, 1.4$\sigma$ and 2.3$\sigma$ outside the
\citet{ih95} tidal radius, using their quoted error of 3.6 arcmin. Based on the number of extratidal giant candidates, they derive
a surprisingly high mass-loss rate: 27\% of the total mass of Carina
per Gyr.

Because their survey covers a relatively small area, contamination by
subdwarfs in their giant region is not a significant problem. We
make a conservative estimate of the number of subdwarfs in their giant
region by using the luminosity function derived in the Appendix for
our smaller giant region: the halo model predicts that 14 subdwarfs
will be found per square degree to V=20.5 ($M\simeq 20.8$).

\subsection{Effect of Photometric Errors}

Carina has quite a low galactic latitude ($b=22$) so there are large
numbers of disk stars in the foreground, and the potential exists for a
significant number to scatter into the giant region of the \mt2/\m51
diagram.
The accuracy of the \citet{srmcarina} photometry is not high.  This is
caused in great part by their short exposure times: their C40
telescope exposure times in $M$ and $T_2$ were almost the same as our 4m
BTC exposure times,  15 times shorter when the difference in
aperture is taken into account. (Their 51 exposures were a factor of
$\sim$3 shorter after correcting for telescope aperture.)  MOPKJG work
more than a magnitude fainter than our survey, to $M=20.8$
($V\simeq$20.5).  In addition, the C100 data suffered from such large
PSF variations that even a quadratic variable PSF in DAOphot was
unable to completely correct them.
 
To deal with these large errors they restrict consideration to
stars with errors in all filters less than 0.10 mag. Thus it is
possible for stars with color errors as high as 0.14 mag in \m51\ or \mt2\
to be included in their sample. In Figure \ref{srmbal} we showed that at
latitudes of 30--40 degrees, significantly higher than the latitude of
Carina, almost 200 dwarf stars per square degree are scattered into
the MOKP giant box for color errors of 0.10 mag. 
Thus the additional cut using the Carina giant branch is very important.

MOPKJG estimate the contribution of ``background'' objects by
successively displacing the Carina giant branch region to {\it brighter}
magnitudes, and counting how many stars which satisfy their \m51/\mt2\
cut are found in the shifted giant branch region. Their decision to
only shift the box to brighter magnitudes is forced upon them by the
serious incompleteness of their photometry at magnitudes fainter than
M=20.8. This is unfortunate, because the number of interlopers in the
giant region 
increases sharply at fainter
magnitudes, so their background estimates will be systematically too
small. 

\begin{figure}
\includegraphics[scale=0.55,angle=270]{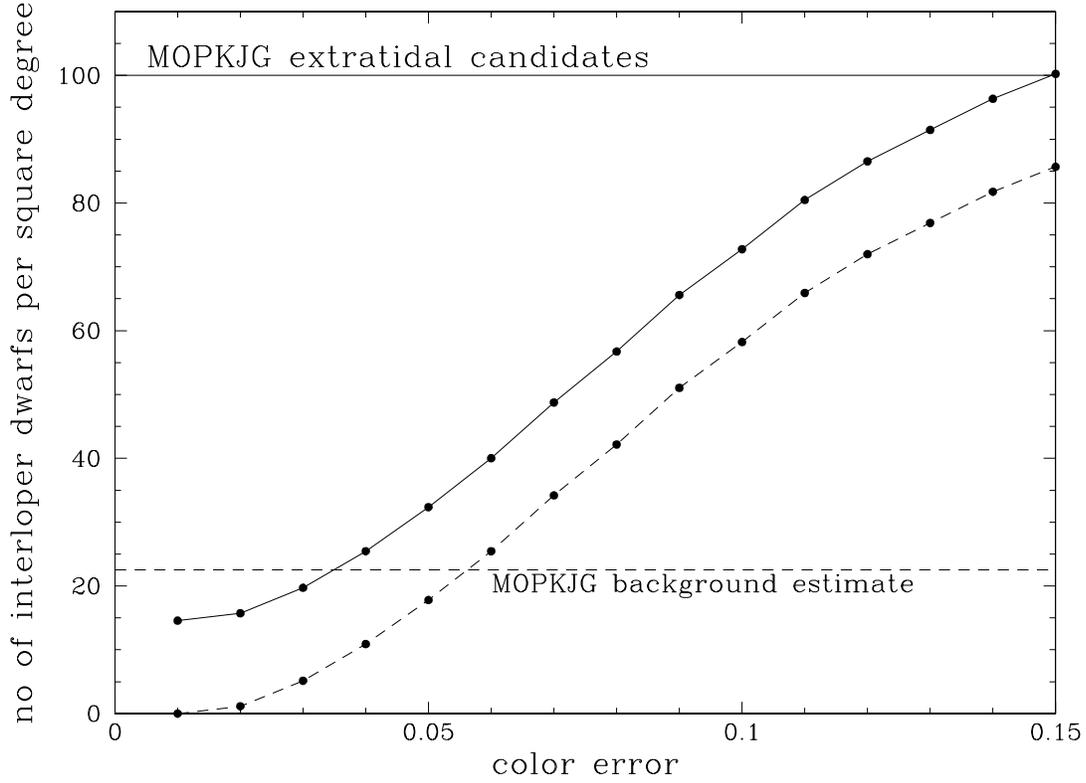}
        
\caption{Our prediction of number of foreground dwarfs (dashed curve)
and foreground dwarfs including metal-poor subdwarfs (solid curve) that
will be scattered into the MOPKJG giant selection regions in the
\mt2/\m51 and \mt2/$M$ diagrams, as a function of photometric error,
for M magnitudes brighter than 20.8. MOPKJG's estimated background
level and number of extratidal giants per square degree are also
shown with straight lines, dashed and solid respectively. 
\label{carinadwarfs}}
\end{figure}

The magnitude range of interest for Carina giants is $M_0$=17.8--20.8.
Here, the majority of the C100 photometry has errors ranging
up to 0.15 mag in $M$ and 51 and 0.25 mag in $T_2$. The C40 data has larger
errors: up to 0.40 mag in 51, 0.25 mag in $T_2$ and 0.3 mag in $M$. With an
error cut of 0.10 mag per filter and a significant number of stars
having errors in each filter close to this number, color errors of up to
0.14 mag are likely.  We have extended our simulations
of the likely number of foreground dwarfs that will be moved into the
MOKP giant selection box in the \m51/\mt2\ diagram by adding the same
selection as was used by MOPKJG (the CMD bounding box outlined in
their Figure 8) to isolate stars at the right color and magnitude to
be on Carina's giant branch.  Figure \ref{carinadwarfs} shows our
results. It can be seen that with color errors in the range of 0.05 to
0.10 mag, dwarf contamination is significantly higher than that
claimed by MOPKJG, whose largest extratidal background
estimate was 22 stars per square degree.  Even the extra cut using the
position of Carina's red giant branch would not rid the extratidal
giant sample of many interlopers.

In summary, more spectroscopic follow-up is needed to confirm the claim
of \citet{srmcarina} that Carina is losing 27\% of its mass per Gyr.
With photometric errors of 0.10 mag, the typical star
in the giant region will be a photometric error. Brighter stars, such
as the three observed spectroscopically,
will have smaller photometric errors and will in general be
giants if they are in the giant region. 
However, only one of these is more than 2$\sigma$ away from the
\citet{ih95} tidal radius. A larger spectroscopic sample is needed.
The discovery of one
extratidal bright giant tells us nothing about the discovery
fraction of fainter giants.

\section{Summary}

Washington photometry, with the addition of the 51 filter, is a
powerful method of finding distant halo giants. 
We find that at the faintest magnitudes probed by our survey, our
giant selection efficiency falls, and we make a careful examination of
possible problems with the Washington giant classification to improve
future surveys.
We show that accurate photometry (color errors less than 0.03 mag) is
essential to cut down on the scattering of the numerous foreground dwarfs
into the giant region in the \m51/\mt2\ color-color plot. 
We have made estimates
of the contamination of giant candidates by very metal-poor halo
dwarfs by deriving a new luminosity function specifically for the
subdwarfs which mimic halo giants in the \m51/\mt2\ diagram, and using
it in a simple halo model.  

We used our results on
luminosity discrimination with the Washington system to examine two
recent surveys: our own survey of 12.7 square degrees at high galactic
latitudes \citep{spag1,spag2}, and the study of \citet{srmcarina} of
fields centered on the Carina dwarf spheroidal galaxy, which finds
over 100 Carina-associated giant candidates outside its tidal radius. 

For surveys such as our own of the halo field, very metal-poor halo
dwarfs are the major source of contamination of candidate giant
samples, and we propose using a smaller region of the \m51/\mt2\
diagram to optimise spectroscopic follow-up at the faint magnitudes
where the subdwarfs dominate.  The survey of \citet{srmcarina} has
spectroscopic follow-up only for three of the brightest ``extra-tidal''
giant candidates, and we show that their photometric errors, which
will increase as stars become fainter, can have a very serious effect
on their giant classification.

In summary, spectroscopic follow-up of giant stars identified using the
Washington system is vital. If one wishes to make statistical
corrections for the number of bogus giants caused by photometric
errors, it is important to quantify accurately the photometric errors
in the data, both in terms of the size of the average error and the
shape of the distribution.

\acknowledgments

We would like to thank the referee, Bruce Carney, for helpful comments.
This work was supported by NSF grants AST 96-19490 to HLM, AST
95-28367, AST96-19632 and AST98-20608 to MM, and AST 96-19524 to EWO.

\vskip 2cm

\appendix{\bf APPENDIX: Luminosity Function for Halo Dwarfs with
[Fe/H] $<$ -2.0}

We have derived a luminosity function (LF) for halo G and K dwarfs with
[Fe/H] $<$ --2.0 using LFs for four metal-poor globular
clusters  derived from HST WFPC2 color-magnitude diagrams by
\citet{piotto}, and the splendid $V$,\vi\ color-magnitude diagram of
NGC6397 of \citet{king98}, transformed to the standard Landolt system
using the transformations of \citet{holtz}. 

We need to derive a LF separately for the metal-poor stars -- its 
slope will be different for the
halo as a whole and for its most metal-poor stars. This can be seen in
the compilation of accurate globular cluster LFs of \citet{paresce},
where for $M_I$=5--8, clusters such as NGC6397 have relatively more
stars at the faint end than metal-richer clusters such as NGC6752 and
NGC104. Also, since we are interested in the luminosity distribution
at a given color, we need to use the color-magnitude diagram of
a metal-poor cluster to decide which absolute magnitudes appear in our
\mt2 color range for stars with \fe $<$ --2.0, where our ability to
discriminate them via \m51 begins to fail. 

\begin{figure}[h]
\includegraphics[scale=0.5,angle=270]{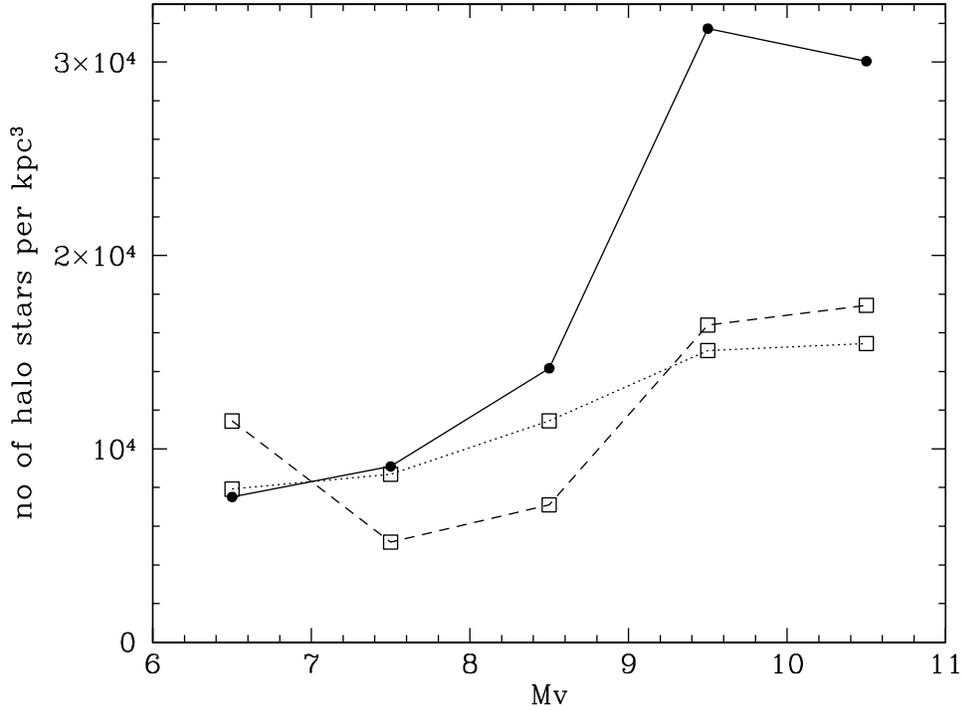}
        
\caption{Comparison of halo field LF of \citet{bc86} (dashed line)
with the average of the accurate LF of NGC 6752 \citep[dotted
line]{paresce} and the average of the LFs of four metal-poor globulars
from \citet{piotto} (solid line). The cluster LFs were normalized to
the field LF for $M_V$=6--8.
\label{lfs}}
\end{figure}

Although it would be more direct to use a subsample of the
\citet{bc86} field star data with \fe $<-2.0$ to derive this LF, the
\citet{bc86} sample is too small. It is based on only 51 stars in our
magnitude range of interest ($M_V$=6--9), and is particularly poorly
sampled at the bright end.  Thus we use accurate LFs derived from HST
photometry of metal-poor globular clusters, where there are sufficient
stars to remove concerns about Poisson counting errors.  We first
compare the LF from a cluster with a similar \fe\ to the mean of the
halo field \citep[][NGC 6752]{paresce} with the \citet{bc86} halo
field LF to check for cluster-field differences, as halo field stars
may not have formed under identical conditions to globular clusters.
(The cluster was observed near its half-light radius where mass
segregarion effects are minimal.) It can be seen in Figure \ref{lfs}
that these two LFs are very similar.

We then proceed to construct our metal-poor halo LF using the clusters
M15, M30, M92 and NGC6397 (whose [Fe/H] values range from
--1.9 to --2.2, Zinn 1985). \citet{piotto} 
compare the LFs of their four globular
clusters and note that three (M15, M30 and M92) are in good agreement,
while the LF of NGC 6397 has fewer stars for $M_V>7$ than the other
clusters. They calculate multi-mass King models in order to check whether
corrections are needed to convert their local LFs to global ones, and
conclude that any corrections  will not exceed one or two tenths in the
logarithm. Thus, since \citet{piotto} were unable to relate the only
difference in the LF slopes to mass segregation effects, we choose to
simply average the four LFs.

Figure \ref{lfs} shows this LF, compared to that of \citet{bc86} for
halo field stars, calculated with a ``discovery fraction'' of 0.5
\citep{hlm93}. The cluster LFs were normalized to agree with the
\citet{bc86} LF for $M_V$=6--8.  As expected, the metal-poor LF has
more stars at the faint end than the \citet{bc86} LF and the NGC 6752
one.

With this luminosity function,
it is possible to predict numbers of K
dwarfs with \fe $<$ --2.0 in the color range corresponding to our giant
box, as a function of magnitude and position in the Galaxy. For
relatively bright stars ($V<19$) we only probe out to distances of 3
kpc at most. As in Paper I, we used the metallicity distribution of
\citet{sean91} to determine that 31\% of halo dwarfs have \fe
$<$ --2.0. The rare subdwarfs with \fe $\leq$--3.0, such as G160-30,
are of particular concern to us because it requires higher signal-to-noise
in the follow-up spectrum to determine whether the star is a dwarf or
giant. We used the metallicity distribution of \citet{beers99}, which
is  complete for metallicities below --2.0, and contains
4754 stars in total, to determine that 3\% of all halo dwarfs have \fe\
$\leq$ --3.0. (Within the errors, this is consistent with the result
of \citet{sean91}.)

\begin{figure}[h]
\includegraphics[scale=0.5,angle=270]{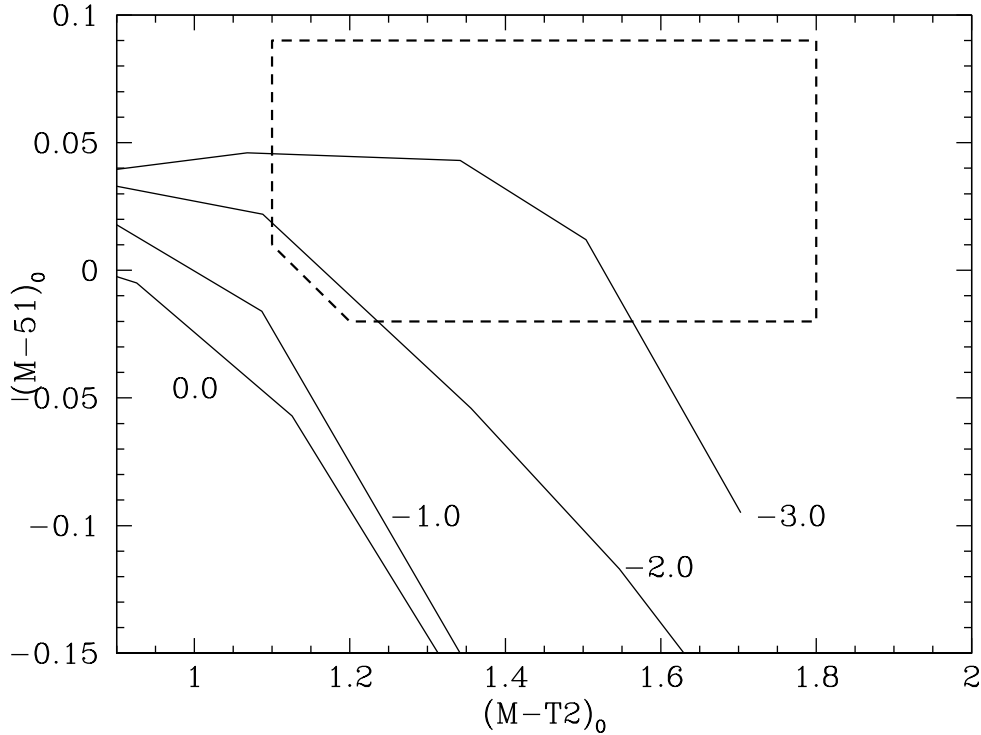}
        
\caption{This figure illustrates the region of our giant selection box
(shown with a dotted line) that overlaps with the synthetic dwarf loci
of \citet{pb94} for \fe = --2.0 and --3.0.
\label{dwarflines}}
\end{figure}

It can be seen in Figure \ref{dwarflines}, which shows the synthetic
colors of \citet{pb94} for dwarfs of different metallicity, that our
giant selection box overlaps the [Fe/H] = --2.0 subdwarfs only at its
blue end (\mt2 = 1.1--1.3) and overlaps the \fe=--3.0 lines for a
larger color interval (\mt2 = 1.1 -- 1.6). In other words, for the
coolest K dwarfs in our color range, the Mg lines are sufficiently
strong that only the most metal-poor (\fe$\leq$--3.0) appear in
our giant selection box, while for the warmer K dwarfs, the lines are
weaker and so stars with \fe $\leq$--2.0  appear in our giant
selection box.  We took this into account in the construction of the
luminosity function for the halo subdwarf model, by scaling the
numbers of dwarfs with bluer colors (brighter absolute magnitudes)
by 31\% and those with redder colors (fainter absolute magnitudes) by
3\%.

Table \ref{lftable} gives the luminosity function for metal-poor subdwarfs
derived from the clusters in \citet{piotto}, and the luminosity
function specifically calculated to match the number of metal-poor
dwarfs that will appear in our giant box.

\begin{deluxetable}{llrr}
\tablewidth{0pt}
\tablecaption{Luminosity functions for metal-poor dwarfs}
\tablehead{
\colhead{$M_V$ range} & \colhead{\mt2\ range} & \colhead{Metal-poor LF} &
\colhead{``Giant interloper'' LF} \\
& & \colhead{(stars per kpc$^3$)} &  \colhead{(stars per kpc$^3$)} 
}

\startdata

6.0--6.5 &0.93--1.05 &3245  & 0  \\
6.5--7.0 &1.05--1.17 &4268  &662 \\
7.0--7.5 &1.17--1.30 &4073  &1262 \\
7.5--8.0 &1.30--1.43 &5014  &150 \\
8.0--8.5 &1.43--1.60 &6036  &181\\
8.5--9.0 &1.60--1.76 &8130  &0 \\
9.0--9.5 &1.76--1.83 &15497  &0 \\

\enddata

\label{lftable}
\end{deluxetable}
\afterpage{\clearpage}





\clearpage









\end{document}